# Radio-Optical Orientation of Elliptical Galaxies

Heinz J. Andernach

*INSA, ESA–IUE Observatory, Apdo. 50727, E-28080 Madrid, Spain*

**Abstract.** Present knowledge of the distribution of the angle between major optical and radio axes of early-type galaxies is reviewed. While brightest galaxies in Abell clusters show a bimodal distribution of this angle, with both parallel and vertical alignment, ellipticals in general only show a weak preference for vertical alignment. Current plans are outlined to homogenize and largely increase these samples using literature data together with digitized optical Sky Surveys and the ongoing new radio surveys. The results will shed light on both the dynamics of nearby ellipticals, as well as on a possible link between nearby cD galaxies and powerful high-redshift radio galaxies.

## 1. Introduction

Powerful radio galaxies at high redshifts ($z \gtrsim 0.5$) tend to show a strong alignment of their radio and optical major axes (Chambers et al. 1987; McCarthy et al. 1987). The observations support two models for the origin of this effect: jet-induced star formation (Dunlop 1995) and scattering by dust of the light of a central AGN (Spinrad 1995). It has been suggested that these objects may be the precursors of dominant cD-like galaxies in clusters at low z (e.g., Djorgovski et al. 1987; Spinrad et al. 1995). If so, then the brightest galaxies of low-z clusters might show signs of radio-optical alignment as well. Even though the **general** population of nearby elliptical radio galaxies favors a radio axis closer to the optical minor axis (*vertical* alignment) a small population of aligned radio galaxies was found among cD-like galaxies (Andernach et al. 1993). This suggests that the environment may influence the relative orientation of optical and radio axes. I describe plans to assess this question by constructing larger samples.

## 2. Brightest Cluster Galaxies

Based on a search of the literature Andernach et al. (1993) constructed a sample of 110 brightest members in Abell clusters with satisfactory data on the *inner* radio position angle and the *outer* optical position angle. The distribution of the radio-optical difference angle $\phi$ (Fig. 1a) reveals two distinct populations. The majority of sources have $\phi > 45°$ with a broad peak near 90°. However, some 20% of the objects show $\phi < 15°$ which is about twice the fraction seen in samples of ellipticals *not* selected for cluster membership (cf. Sect. 3). The alignment tendency does not seem to depend significantly on richness or BM-type of the clusters, nor on redshift (median z is 0.06) or largest linear size ($\text{LLS}_{med} = 200\ h_{50}^{-1}$ kpc) of the sources. I have now increased the sample to 155 objects and included radio fluxes at various frequencies. It turns out that the histogram of $\phi$ is not sensitive to radio luminosity either, but indeed changes with the radio spectral index, in the sense that both the bimodality and the fraction of aligned sources is highest for the steeper spectrum sources ($\alpha > 0.78$, where $S \sim \nu^{-\alpha}$). This is interesting since the steep radio spectrum (together with optical faintness) is the operational criterion by which the aligned high-z radio galaxies are being found. It is true that the aligned, high-z radio galaxies have radio luminosities which are 2–5 orders of magnitude greater than those in the low-z sample of brightest cluster galaxies, and thus the alignment is likely to have a different physical origin. However, the relation between alignment and steep radio spectrum is not understood, and a more detailed study of the low-z objects may provide clues to the origin of the alignment at both low and high redshifts.

## 3. Ellipticals in General

Radio-optical orientations based on various samples of ellipticals had been discussed in eleven papers until Sansom et al. (1987) attempted a synthesis of these samples. They note that the *inner* and *outer* optical p.a. is sometimes different for a given galaxy due to its isophotal twisting. Nevertheless, they merged their sample (based on *inner* optical and radio isophotes) with the previous ones, most of which were based on *outer* optical and radio isophotes. The combined sample of 197 objects showed a relatively flat distribution of $\phi$ with only a weak preference for vertical alignment.

To further increase the sample, I used data from three further papers, ignoring the differences between the authors as regards the exact definition of radio and optical axes. Condon et al. (1991) provide 59 E/S0 galaxies not contained in Sansom et al. (1987). Andreasyan (1984) contains data for an additional 57 and Sadler et al. (1987) for another six objects. The full sample of 319 objects (Fig. 1b) shows only a marginal preference for vertical alignment, in contrast to some of the subsamples (Condon et al. 1991 or Palimaka et al. 1979) which suggested this preference to be more pronounced. By comparison with Fig. 1a it may seem that the cluster environment causes the aligned fraction of objects to increase. However, Sansom et al. (1987) found that the 15 clearly "isolated" galaxies in their sample show a pronounced bimodality of the $\phi$–distribution similar to Fig. 1a.

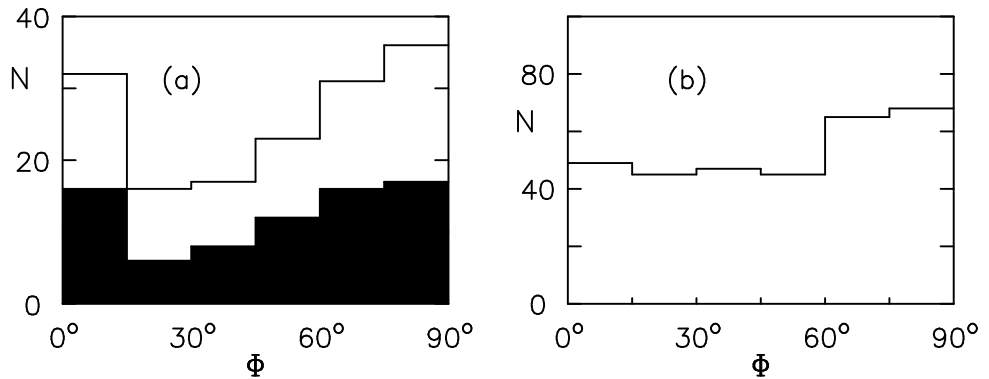

Fig. 1. (a) Radio-optical position angle difference for 155 brightest members of Abell clusters. The black area marks good quality data with $\Delta\phi \lesssim 10°$. (b) Same as (a) but for the full sample of 319 radio ellipticals (see Sect. 3).

## 4. Theoretical Implications

Two papers have so far shown how this analysis can improve our theoretical understanding of radio ellipticals. Andreasyan (1984) has looked at the radio morphology of objects in his sample and found that the distribution of $\phi$ depends critically on the ratio R=(length/width) of the overall radio structure, in the sense that sources with R<2.5 clearly prefer parallel alignment ($\phi \approx 0°$) and those with R>2.5 prefer a vertical one ($\phi \approx 90°$). The author interprets this as evidence for a general dipole magnetic field along the *minor* (rotational?) axis of the galaxy. In sources with R<2.5 the magnetic field energy dominates over the kinetic energy of the radio plasmons, thus the relativistic electrons get trapped in a toroidal region causing radio lobes along the *major* optical axis. In sources with R>2.5 the plasmons reach great distances from the parent galaxy along the dipole (optical minor) axis. The problem with this interpretation is the lack of evidence for a general rotation of ellipticals about their optical minor axes.

Sansom et al. (1987) performed simulations in which the radio plasmons are ejected along certain directions with respect to the principal axes of triaxial ellipsoids, including the limiting cases of prolate and oblate spheroids. They found that a distribution of $\phi$ like in Fig. 1b can best be explained by a prolate spheroid with the radio axis on average along the optical minor axis, but with a spread of some $\pm 30°$. A bimodal histogram like in Fig. 1a requires an ejection along the intermediate axis of triaxial ellipsoids. The authors assumed the ellipsoid to be of intrinsic type E2, but a more thorough analysis will have to take into account a realistic distribution of ellipticities as observed in nature.

## 5. Work In Progress

It is obvious that the data contained in Fig. 1b are drawn from studies too heterogeneous to allow a meaningful comparison with Fig. 1a which was prepared more carefully. Moreover, the sample in Fig. 1b contains some brightest cluster members and for many objects the cluster environment has never been checked. It is thus necessary to establish a consistent data set for all objects, including all parameters which are suspected to influence the radio-optical orientation.

In an attempt to further increase the sample size I cross-identified the currently largest catalogues of radio sources and galaxies which cover the entire sky. These are the 87GB and PMN with a total of ∼100,000 radio sources (cf. Griffith et al. 1995 and references therein) and the "Lyon-Meudon Extragalactic Database" (LEDA, Paturel et al. 1995) with likewise ∼100,000 galaxies. Allowing for the positional errors and sizes of both radio sources and galaxies, I found that for ∼1400 different radio sources there are ∼1600 galaxies sufficiently close to serve as good candidate optical counterparts. Of these galaxies, 24% are of early type (T<0) and 28% are late-type (T>0) while the rest is of unknown type. After verifying each identification the relevant radio-optical data will be collected for these objects in appropriate compilations and databases, in ongoing radio surveys (e.g. NVSS and FIRST at the VLA) and in the digitized sky surveys.

To eventually identify those parameters which most influence the radio-optical orientation one needs to compile e.g. *inner and outer* optical position angle, isophote twist and ellipticies (see e.g. Colina & de Juan 1995), the morphological type, optical magnitudes on a coherent scale, the radio luminosity and spectrum, the radio morphology and core flux fraction, and the galaxy environment (interactions, clusters, etc). For these and many other projects it would be highly desirable to populate existing databases with relevant data and with pointers to published radio maps or even establish a self-standing database of radio sources. Towards this end the author has collected a vast amount of published catalogues and data tables on radio sources yet unavailable from data centers.

## Discussion

*G. Djorgovski*: At high z the alignment effect seems to be connected only with the most powerful, 3C-type sources. Studies of PSR sources by Peacock and Dunlop, and B3 sources by Thompson, Vigotti, Grueff and myself have shown that going down in power by only a factor of 5–10 from 3C's makes the alignment practically disappear. Alignments seen at low z's may be entirely different in origin.

*Andernach*: Yes, this is also suggested by the fact that in our low-z sample we see no increase of the (already small) fraction of aligned sources with radio power.